\documentclass[twocolumn,prl,aps,superscriptaddress,english,floatfix]{revtex4-1}
\usepackage{amsmath}
\usepackage{amssymb}
\usepackage{amsfonts}
\usepackage[pdftex]{graphicx}
\usepackage{subfigure} 
\usepackage[english]{babel}
\usepackage{braket}
\usepackage{color}
\usepackage{mathtools}
\usepackage{threeparttable}
\usepackage[colorlinks,linkcolor=blue,anchorcolor=blue,citecolor=blue,urlcolor=blue]{hyperref}
\usepackage{babel}
\makeatother

\begin{document}

\preprint{APS/123-QED}

\title{Confining light in all-dielectric anisotropic metamaterial particles for nano-scale nonlinear optics}

\author{Saman Jahani}%
\author{Joong Hwan Bahng}%
\author{Arkadev Roy}%
\affiliation{%
Department of Electrical Engineering, California Institute of Technology, Pasadena, CA 91125, USA.
}%
\author{Nicholas Kotov}%
\affiliation{
Department of Chemical Engineering, University of Michigan, Ann Arbor, MI 48109, USA.
}%
\author{Alireza Marandi}%
\email{marandi@caltech.edu}
\affiliation{%
Department of Electrical Engineering, California Institute of Technology, Pasadena, CA 91125, USA.
}%

\date{\today}

\begin{abstract}
High-index dielectrics can confine light into nano-scale leading to enhanced nonlinear response. However, increased momentum in these media can deteriorate the overlap between different harmonics which hinders efficient nonlinear interaction in wavelength-scale resonators in the absence of momentum matching. Here, we propose an alternative approach for light confinement in anisotropic particles. The extra degree of freedom in anisotropic media allows us to control the evanescent waves near the center and the radial momentum away from the center, independently. 
This can lead to a strong light confinement as well as an excellent field overlap between different harmonics which is ideal for nonlinear wavelength conversion. 
Controlling the evanescent fields can also help to surpass the constrains on the radiation bandwidth of isotropic dielectric antennas. This can improve the light coupling into these particles, which is crucial for nano-scale nonlinear optics.
We estimate the second-harmonic generation efficiency as well as optical parametric oscillation threshold in these particles to show the strong nonlinear response in these particles even away from the center of resonances. Our approach is promising to be realized experimentally and can be used for many applications, such as large-scale parallel sensing and computing.

\end{abstract}

\maketitle

\begin{figure*}
\centering
\begin{tabular}{cc}

\includegraphics[width=17cm]{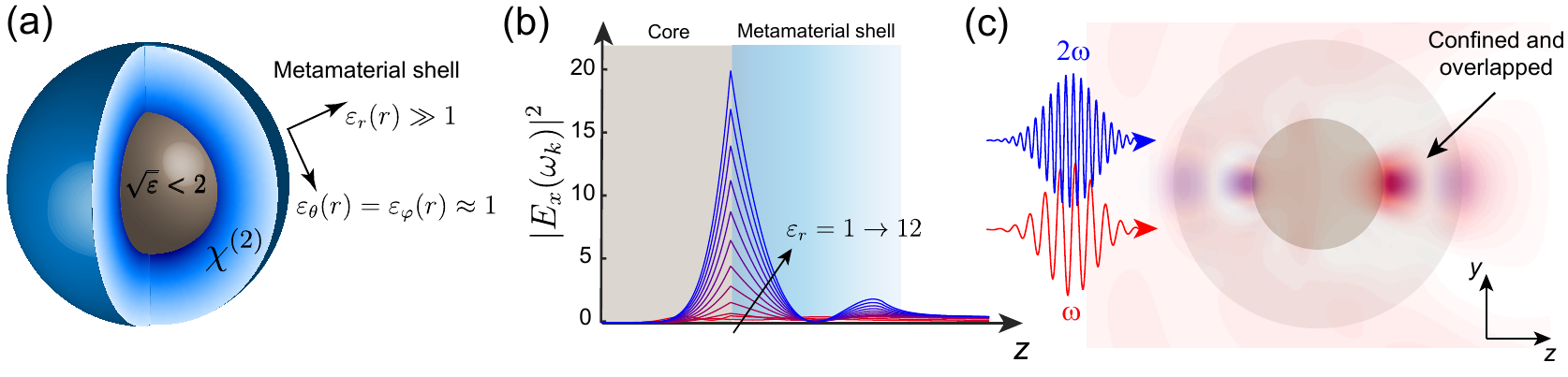}

\end{tabular}
\caption{{\bf Confining light in anisotropic metamaterial particles to enhance nonlinear interaction}. (a) Schematic representation of a low-index ($\sqrt{\varepsilon}<2$) particle with metamaterial shell. The radial anisotropy of the shell, with an optical axis in the $r$ direction, offers an extra degree of freedom to engineer the electric Mie modes of the particle. $\varepsilon_{\theta}$ and $\varepsilon_{\varphi}$ control the momentum while $\varepsilon_r$ can control the order of spherical waves. (b) The electric field distribution of the 5$^{th}$ electric mode at resonance as a function of $\varepsilon_r$ in the shell while $\varepsilon_{\bot}=\varepsilon_{\theta}=\varepsilon_{\varphi}=1$. The core is isotropic $\varepsilon=2.2$ with a radius of $R_1=0.5$~$\mu$m, and the shell radius is $R_2=1.1$~$\mu$m. The shell has a graded-index profile such that $\varepsilon_r(R_2)=1$. Increasing the anisotropy enhances the field at the core/shell interface. Similar effect can be seen for other electric modes as well. (c) The anisotropy of the shell can enhance the fields at second harmonic at the core/shell interface as well. The field enhancement and strong overlap can significantly enhance the nonlinear interaction for efficient second-harmonic generation and optical parametric oscillation processes. 
} 
\label{fig:Anisotropic_particle_Schematic2}
\end{figure*}

Efficient nonlinear light generation requires long range nonlinear interaction and/or strong field enhancement \cite{boyd2019nonlinear}. For massive computing and sensing in mid-infrared, it is desirable to miniaturize nonlinear systems to nano-scale \cite{mcmahon2016fully, wang2018quantum, yesilkoy2019ultrasensitive}.
However, miniaturization of photonic devices to nano-scale not only reduces the interaction length, but also deteriorates the light confinement because of the diffraction limit of light.
Plasmonic and epsilon-near-zero structures can enhance light confinement at nanoscale leading to strong nonlinear response  with limitations due to the optical loss of metals \cite{pu2010nonlinear, zhang2011three, nielsen2017giant, reshef2019nonlinear}. 

Recently, light confinement in all-dielectric high-index nano-structures has emerged as a low loss alternative to enhance the nonlinear response at nano-scale \cite{smirnova2016multipolar, krasnok2018nonlinear, pertsch2020nonlinear, shcherbakov2014enhanced, yang2015nonlinear, gili2016monolithic, camacho2016nonlinear,  marino2019spontaneous, koshelev2020subwavelength, saerens2020engineering}. The high-Q Mie resonances in high-index particles with sub-wavelength sizes can help to confine energy inside the particles which can be beneficial for nonlinear wavelength conversion \cite{jahani2016all,kuznetsov2016optically,baranov2017all}. However, in isotropic media, the momentum increases with increasing the refractive index. This constrains the field overlap especially for higher order high-Q modes. As a result, exploiting higher order modes without a proper momentum matching does not necessarily improve the nonlinear response \cite{gigli2020quasinormal, jahani2020wavelength}. Besides, because of inefficient radiation of high-Q nano-antennas \cite{wheeler1947fundamental, chu1948physical, collin1964evaluation, sievenpiper2011experimental, ziolkowski2003application, li2019beyond}, in/out-coupling in high-index dielectric nano-antennas is weak which degrades the nonlinear conversion efficiency in these particles. 

Light can be confined using low-index particles based on the multi-mode interaction in which due to the low-Q and small momentum of light, multiple modes can spatially and spectrally overlap and form a bright hot-spot which is known as `photonic nanojet' \cite{luk2017refractive,chen2004photonic}. However, the intensity of the hot-spot in simple configurations is directly proportional to the size of the particle, which hinders miniaturization. Besides, the hot-spot is usually formed outside the particle. Hence, it is difficult to construct an overlap between the optical mode and a nonlinear material.

Here, we propose a paradigm shift in light confinement approaches and nonlinear optics at nano-scale using low-index particles surrounded by all-dielectric anisotropic metamaterial shell ($\varepsilon_{r} \gg 1$, $\varepsilon_{\bot}=\varepsilon_{\theta}=\varepsilon_{\varphi}\approx 1$, and $\mu_{ij}=1$) with second-order nonlinearity (Fig.~\ref{fig:Anisotropic_particle_Schematic2}(a)). 
The anisotropy of the shell gives us two degrees of freedom to independently control the radial momentum of light and the penetration of large angular momentum states toward the center. Hence, we can enhance the field intensity without a significant change in the field profile (Fig.~\ref{fig:Anisotropic_particle_Schematic2}(b)). 
This allows us to achieve field confinement as well as strong overlap between harmonics (Fig.~\ref{fig:Anisotropic_particle_Schematic2}(c)) leading to enhanced  wavelength conversion in these particles. We show that the radial anisotropy can also help to convert reactive energy near the center into propagating waves, and as a result, the radiation properties of these particles are remarkably improved. As a result, the efficiency of the coupling of input light and collection of the output light is enhanced. 
We estimate the second-harmonic generation (SHG) efficiency as well as the optical parametric oscillation threshold in these particles. We discuss that the nonlinear response in the proposed particles can be orders of magnitude higher compared to a high-index isotropic particle with similar nonlinear coefficient and Q factors. 

Natural low loss dielectrics have limited anisotropy \cite{jahani2016all}. However, strong anisotropy can be achieved in high contrast dielectric nanostructures \cite{jahani2018controlling, xu1995fabrication, niu2018giant, liu2019direct, palmer2020highly, green2020optical, ermolaev_giant_2021}. Recently, we have experimentally demonstrated particles with anisotropic metamaterial shells \cite{bahng2020mie}. 
The metamaterial shell is composed of dielectric nanowires arranged in spherical form \cite{bahng2015anomalous}. Due to the sub-wavelength feature size of the unit-cell in the shell, Maxwell-Garnett effective medium theory can be applied to model the effective response of the shell \cite{jahani2014transparent}. Since the nanowires are mostly oriented in the radial direction, the nanowires demonstrate an effective spherical anisotropy. Also, as the distance from the center is reduced, the nanowires filling factor reduces while the width of the nanowires is fixed. Hence, the effective response displays a graded-index profile as well. The analytical calculation of the field distribution using a modified Mie theory is in good agreement with the full-wave simulation except near the nanowires which is due to the inhomogeneity of the real structure \cite{SM}. The anisotropy that we have achieved with zinc oxide nanowires in our recent experiment \cite{bahng2020mie} is limited, but it can be enhanced by using higher index nanowires \cite{schuller2007dielectric} or doping the nanowires \cite{riley2016high}.

To understand the light confinement mechanism in these particles, we first look at the wave equations in media with spherical anisotropy. Since, the magnetic modes (TE modes) are not affected by the non-magnetic anisotropy, we only focus on the electric modes (TM modes) here. The wave equation in uniaxial media with optical axis in the $r$ direction can be written as \cite{SM}:
\begin{align}
\label{eqn:wave_eqn}
    -\frac{1}{\varepsilon_{\bot}}\frac{1}{r^2}\frac{\partial}{\partial{r}}\left(r^2\frac{\partial}{\partial{r}}(rE_r)\right) + \frac{1}{\varepsilon_{r}r^2}\Vec{\pmb{L}}^2(rE_r)
=k_0^2(rE_r),
\end{align}
where $\hbar\vec{\pmb{L}}=\frac{\hbar}{i}(\Vec{r}\times \Vec{\nabla})$ is the angular momentum operator with an eigenvalue of $\hbar \sqrt{n(n+1)}$ and $n$ is an integer describing the angular momentum mode number \cite{jackson2007classical}. The first term on the left-hand side of Eq.~\ref{eqn:wave_eqn} corresponds to the radial momentum with an eigenvalue of $\hbar k_r$ which can be expressed as \cite{SM}:
\begin{align}
\label{eq:dispersion}
  \frac{k_r^2}{\varepsilon_{\bot}}+\frac{n(n+1)}{\varepsilon_{r}r^2}=k_0^2.  
\end{align}
The radial component of the electric field in a homogeneous media with spherical anisotropy excited by a plane wave can be written as a superposition of orthogonal modes \cite{SM}:
\begin{align}
     E_r(r,\theta,\varphi) &=\frac{1}{(k_0r)^2}\sum_{n=1}^{\infty}{ c_nz_{n_e}(k_0\sqrt{\varepsilon_{\bot}}r)
     P_n^{(1)}\left(\cos{\theta}\right)e^{\pm i\varphi}}, \\ \nonumber
     {n_e} &=\sqrt{\frac{\varepsilon_\bot}{\varepsilon_r}n(n+1)+\frac{1}{4}}-\frac{1}{2},
\end{align}
where $P_n^{(1)}$ is the associated Legendre polynomial of the first order, $z_{n}$ is one of the Ricatti-Bessel functions or their superposition \cite{bohren2008absorption, qiu2009spherical}, $c_n$ is the amplitude of the $n^{th}$-mode, $k_0=\omega/c$ is the momentum in free-space, $\omega$ is the angular frequency, and $c$ is the speed of light in vacuum.  

\begin{figure}
\centering
\begin{tabular}{cc}

\includegraphics[width=8.5cm]{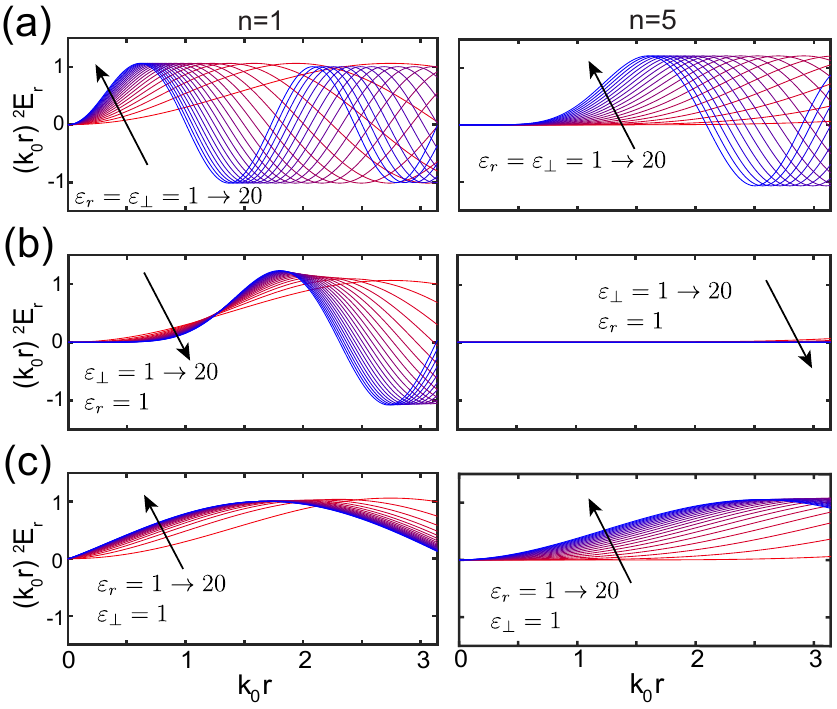}

\end{tabular}
\caption{{\bf Field distribution in an infinite homogeneous media with radial anisotropy}. Normalized electric field distribution for the first (left) and the fifth (right) electric modes as a function of permittivity. (a) Isotropic media. (b) Anisotropic media with $\varepsilon_r=1$. (c) Anisotropic media with $\varepsilon_\bot=\varepsilon_\theta=\varepsilon_\varphi=1$. $\varepsilon_\bot$ controls the momentum of spherical waves while $\varepsilon_\bot/\varepsilon_r$ changes the order of spherical Bessel waves. By increasing $\varepsilon_r$ while $\varepsilon_\bot$ is fixed, we can reduce the order without increasing the momentum. This results in an enhanced field intensity, especially in the sub-wavelength regime ($k_0r \ll 1$).   
} 
\label{fig:Anisotropic_particle_Waves}
\end{figure}

Figure~\ref{fig:Anisotropic_particle_Waves} displays the electric field in media with and without spherical anisotropy. We have plotted only the first and the fifth modes. Without the loss of generality, the same arguments can be applied to other electric modes as well.
By increasing the angular momentum mode number, as seen in Eq.~\ref{eq:dispersion}, the radial momentum reduces, and at some point, it becomes imaginary. This causes the field decays faster when it approaches toward the center, which hampers light concentration with large angular momentum in sub-wavelength regime in isotropic structures. This also causes a weak radiation of generated light in the sub-wavelength regime \cite{jacob2006optical}.  

Increasing the refractive index in isotropic media can compress modes in the radial direction which results in increasing the radial momentum as well as enhancing the penetration of evanescent waves toward the center (Fig.~\ref{fig:Anisotropic_particle_Waves}(a)). 

Although the far-field momentum is independent of $\varepsilon_r$ as seen in Eq.~\ref{eq:dispersion}, increasing $\varepsilon_\bot$ alone does not enhance the field near the center (Fig.~\ref{fig:Anisotropic_particle_Waves}(b)). This is due to the suppression of evanescent waves \cite{jahani2014transparent}. This type of anisotropic media can be utilized to control the total internal reflection and to confine evanescent waves inside an isotropic core \cite{jahani2014transparent,jahani2015breakthroughs, liu2016q, liu2016q2, liu2019direct, ermolaev_giant_2021}. 

\begin{figure}
\centering
\begin{tabular}{cc}

\includegraphics[width=8.5cm]{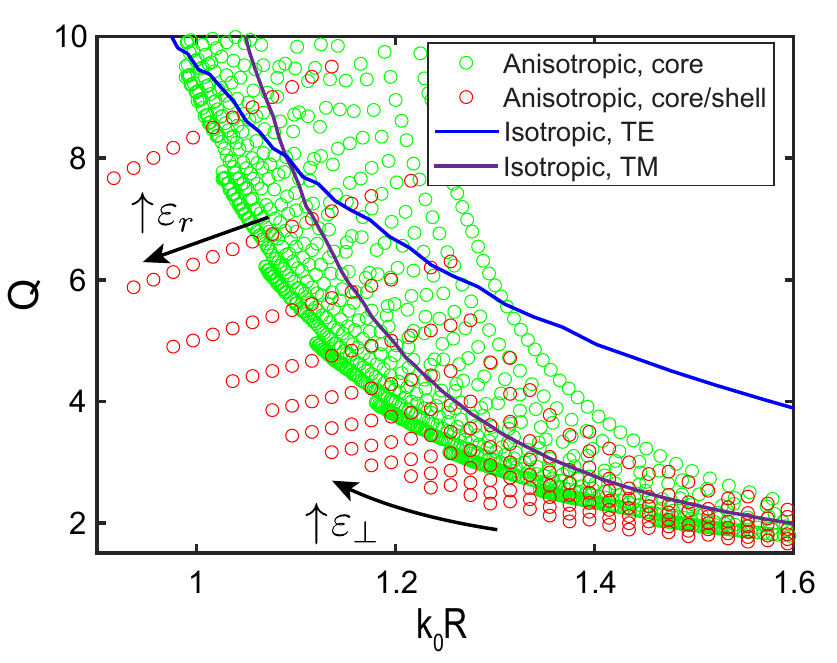}

\end{tabular}
\caption{{\bf Comparing the Q value versus the size in different types of spherical all-dielectric particles ($\varepsilon_{ij}>1$ and $\mu=1$ )}. Each point represents the resonant frequency and the Q of the lowest order mode for a given value of anisotropy with a fixed total size ($R$). The core in core/shell structure is isotropic with $\varepsilon=2.2$ and the size of the particle is the same as that shown in Fig.~\ref{fig:Anisotropic_particle_Schematic2}a.  Note that the magnetic modes are not affected by the dielectric anisotropy. The radial anisotropy can help to surpass the limit on the minimum radiation quality factor of dielectric antennas.
} 
\label{fig:Anisotropic_particle_Q}
\end{figure}

On the other hand, if we increase the anisotropy in the opposite direction, as shown in Fig.~\ref{fig:Anisotropic_particle_Waves}(c), near-field evanescent waves can be enhanced without a significant change in the momentum away from the center. The field enhancement using this approach in subwavelength regime is more substantial than increasing the permittivity in isotropic media (see the Supplementary Materials \cite{SM}) even though the averaged permittivity in the anisotropic media is lower. 
This can lead to a strong conversion of reactive (evanescent) fields near the center into propagating electromagnetic waves even without using hyperbolic structures \cite{,jacob2006optical}. 
As a result, beside the field enhancement, it is expected that the radiation from a particle composed of a material with radial anisotropy to outperform an isotropic dielectric nanoantenna.

\begin{figure*}
\centering
\begin{tabular}{cc}

\includegraphics[width=17cm]{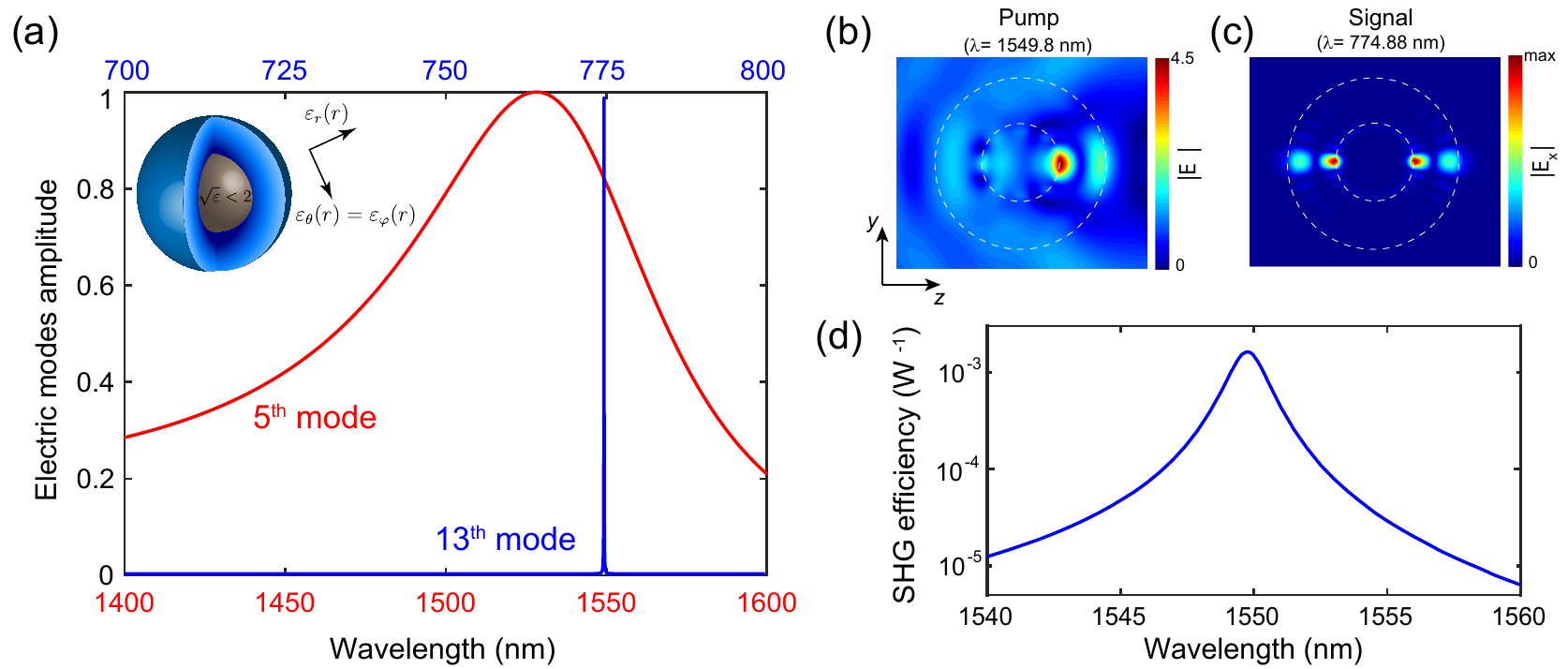}

\end{tabular}
\caption{{\bf Second-harmonic generation in particles with anisotropic metamaterial shell}. (a) Linear response of the modes at the fundamental and second-harmonic frequencies for a low-index particle with anisotropic metamaterial shell. The parameters for the particle are the same as those in Fig.~\ref{fig:Anisotropic_particle_Schematic2}(b) with $\varepsilon_r(R_1)=12$. The normalized scattering amplitude of the 5$^{th}$ (red) and 13$^{th}$ electric modes of the particle. The Q factor for the modes are 25 and 1.6e4, respectively. The second harmonic of the 5$^{th}$ mode coincides with the 13$^{th}$ mode. The contributions of other modes on SHG are negligible because of weak scattering response at the operating wavelengths.
(b) The electric field distribution at the pump wavelength ($\lambda=1549.8$~nm) when the particle is excited by an $x$-polarized plane-wave propagating in the $z$ direction. The electric field amplitude is normalized to the amplitude of the plane-wave. (c) The electric field distribution of the 13$^{th}$ electric mode which resonates at the second-harmonic of the pump excitation. Due to the anisotropy of the shell, the field is enhanced at the interface between the core and the shell. (d) Second-harmonic generation efficiency as a function of the pump wavelength. All the contributing modes at the pump and the signal wavelengths are taken into account. The efficiency boosts as the second-harmonic wavelength approaches the resonance of the the 13$^{th}$ electric mode.
} 
\label{fig:Anisotropic_particle_SHG}
\end{figure*}

To describe the radiation properties of an anisotropic nano-antenna, we have calculated the Q values in anisotropic spherical particles compared to the isotropic case (Fig.~\ref{fig:Anisotropic_particle_Q}). The Q of an antenna is defined by the power radiated by the antenna and the reactive energy stored in it ($Q=\omega W_{\rm stored}/P_{\rm radiated}$), and it specifies the inherent limitation of the physical size of an antenna on its performance has been explored in the classical works by Chu, Wheeler, and others \cite{wheeler1947fundamental, chu1948physical, collin1964evaluation, sievenpiper2011experimental, ziolkowski2003application, li2019beyond}. 
Although increasing the Q is desirable for field enhancement and increasing light-matter interaction in a resonator \cite{vahala2003optical}, it causes an increase in reactive power resulting inefficient coupling of light from and into the far-field. In bulk Fabry-Perot or whispering-gallery-mode resonators, efficient coupling is still achievable by evanescent coupling or impedance matching of the input port.
However, in nano-scale resonators in which multipolar modes can only be excited from the far field, the radiation properties of the resonator play significant roles for light-matter interactions. Figure~\ref{fig:Anisotropic_particle_Q} displays the Q factor of the first electric and the first magnetic modes in isotropic and anisotropic particles. The Q factor in core/shell anisotropic structures can be reduced and approach the Chu limit of dielectric antennas \cite{schuller2009general}. The same approach can also 
be used to improve the radiation of dielectric resonant antennas in the microwave regime where strong anisotropy is more accessible \cite{catrysse2011transverse}.
A similar argument can be applied to the higher order electric modes.

A particle composed of a low-index core and an anisotropic shell (Fig.~\ref{fig:Anisotropic_particle_Schematic2}(a)) can enhance and confine light at the core/shell interface. The evanescent field enhancement because of the anisotropy of the shell as well as the field enhancement in low-index core because of the continuity of the normal component of the displacement current lead to generation of a hot-spot at the boundary for electric modes ( Fig.~\ref{fig:Anisotropic_particle_Schematic2}(b)).




\begin{figure}
\centering
\begin{tabular}{cc}

\includegraphics[width=8.5cm]{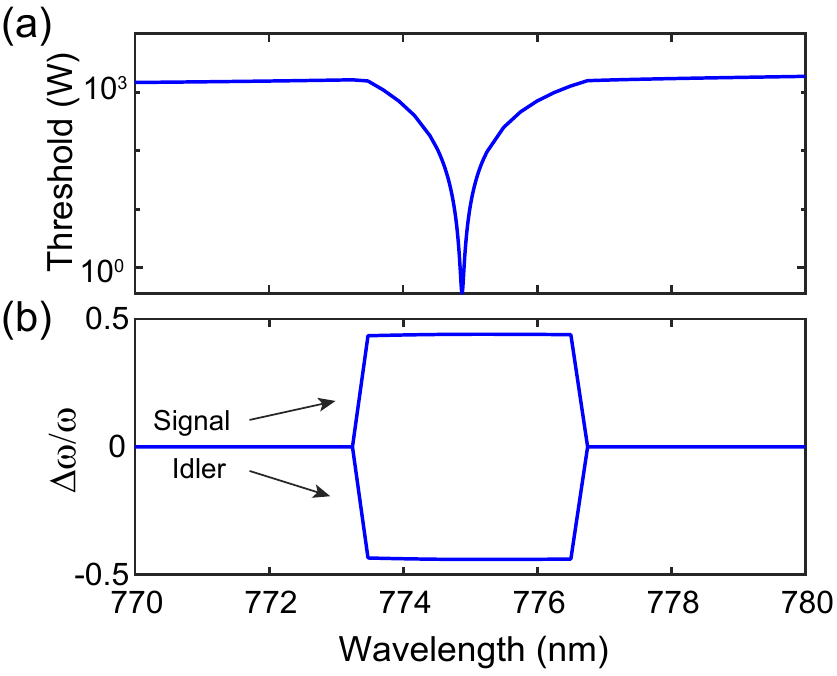}

\end{tabular}
\caption{{\bf Optical parametric oscillation in particles with anisotropic metamaterial shell}. The structure is the same as shown in Fig.\ref{fig:Anisotropic_particle_Schematic2}. (a) Oscillation threshold and (b) signal and idler separation as a function of the pump wavelength. All the contributing modes at the pump and the signal wavelengths are taken into account. The oscillation threshold drops remarkably as the pump wavelength approaches the resonance of the the 13$^{th}$ electric mode. Because of the detuning of the resonant frequency of signal/idler modes from the fundamental harmonic and nonlinear interactions between multiple modes a phase-transition from degenerate to non-degenerate case can happen. 
} 
\label{fig:Anisotropic_particle_OPO_ND}
\end{figure}

Since all the excited electric modes are confined at the core/shell interface, there is a strong spatial overlap between different harmonics at the hot-spot. This can lead to enhanced nonlinear wavelength conversion in these particles. We consider an extreme anisotropy for the shell (inset of Fig.~\ref{fig:Anisotropic_particle_SHG}(a)) to emphasize the role of anisotropy for light confinement and wavelength conversion. The scattering coefficients for the electric and magnetic modes are displayed in the Supplemental Material \cite{SM}. As expected, the magnetic modes are not altered by the shell since they are TE modes. However, the electric modes are significantly affected by the anisotropic shell leading to a field enhancement. 

We choose the fundamental harmonic to resonate at the 5$^{th}$ electric mode. The second harmonic spectrally overlaps with the 13$^{th}$ electric mode. The scattering coefficient for these modes are plotted in Fig.~\ref{fig:Anisotropic_particle_SHG}(a). The scattering coefficient of other modes are illustrated in the Supplementary Materials \cite{SM}. There is a good spectral overlap between the second-harmonic of the 5$^{th}$ with the 13$^{th}$ modes. Hence, they can be employed for the SHG and optical parametric oscillation processes. 
We first look at the SHG process in these particles. We have assumed that the core has no nonlinearity and the shell has a second-order nonlinearity with $\chi^{(2)}=200$~pm/V. We excite the particle with a plane wave which excites multiple modes of the particle (Fig.~\ref{fig:Anisotropic_particle_SHG}(b)) at fundamental harmonic. At second-harmonic, multiple modes can resonate as well. However, since the 13$^{th}$ electric mode has the highest Q around the second harmonic (Fig.~\ref{fig:Anisotropic_particle_SHG}(c)), most of the pump power is converted to this mode if the detuning from the resonant frequency is negligible \cite{jahani2020wavelength, bahng2020mie}. The calculated SHG efficiency considering all the contributing modes is plotted in  Fig.~\ref{fig:Anisotropic_particle_SHG}(d). The SHG efficiency can reach up to $2\times 10^{-3}$~W$^{-1}$ near the resonance. The highest measured SHG efficiency in single dielectric particles is $\sim 10^{-5}$~W$^{-1}$ \cite{camacho2016nonlinear, koshelev2020subwavelength}. It is noteworthy that using higher order modes in isotropic high-index dielectrics does not considerably improve the SHG efficiency without leveraging the phase matching \cite{gigli2020quasinormal}.

We have recently proposed the possibility of parametric oscillation in wavelength-scale resonators \cite{jahani2020wavelength}. Optical parametric oscillators (OPOs) can generate entangled photon pairs and squeezed vacuum states below the oscillation threshold \cite{wu1986generation, morin2014remote, nehra2019state}, while above the threshold at which the gain exceeds loss, they can generate  mid-IR frequency combs which can be used for many applications, such as metrology, spectroscopy, and computation at degeneracy \cite{,eckardt1991optical, muraviev2018massively, marandi2014network}. As we miniaturize a conventional resonator, the nonlinear gain is reduced and field overlap deteriorates if there is no phase matching. As a result, it becomes extremely difficult to surpass the threshold. Since the SHG efficiency is strikingly high in the anisotropic particles introduced here, it is expected to achieve a low oscillation threshold in these particles as well.

Figure~\ref{fig:Anisotropic_particle_OPO_ND}(a) displays the OPO threshold of the first oscillating mode. The minimum threshold is around 0.37 W which happens when the pump overlaps with the 13$^{th}$ electric mode. This threshold is one order of magnitude lower than an isotropic particle with similar values for Q  and nonlinearity \cite{jahani2020wavelength}. This improvement is due to the field enhancement and localization which is not achievable in isotropic particles. Due to the detuning of the resonant frequency of the signal/idler from the fundamental harmonic, the signal and idler separation is large. However, the nonlinear interactions between them can lead to a phase transition from non-degenerate to degenerate case \cite{jahani2020wavelength, arkadev2021spectral}. By engineering the resonant frequency of the modes and reducing the detuning, the OPO threshold can be reduced further.  

It is noteworthy that even away from the center of the resonance of the 13$^{th}$ electric mode, the nonlinear response is still significant compared to an isotropic particle \cite{jahani2020wavelength}. Especially for OPO case, if we are in the low-Q regime, we can compress the pump into an ultra short pulse which can lead to a considerable reduction in the threshold.

In summary, we have proposed a light confinement strategy using particles with spherical anisotropic shell. We showed that in media with spherical anisotropy, the evanescent fields can be enhanced in the sub-wavelength regime without a significant change in the field profile. This field enhancement in sub-wavelength regime, is even stronger than the field enhancement in high-index isotropic media. This allows to confine light in particles with a low-index core and an anisotropic metamaterial shell and localize modes at the core/shell interface for all the electric modes. Controlling the evanescent waves in the sub-wavelength regime can also improve the radiation properties of the nanoantennas which is essential for the efficient excitation and the collection of generated light. 
Our approach also suggests a strong field overlap between different harmonics. We have shown that if the shell is composed of a material with second-order nonlinearity, we can enhance the SHG efficiency and reduce the threshold of OPOs. 
Particles with anisotropic shell are achievable at optical frequencies \cite{bahng2020mie,palmer2020highly, liu2019direct}, and they can open opportunities for exploring nonlinear optics at nano-scale. Even though we have focused on spherical particles, the same concept can be applied to cylindrical Mie resonators which are more amenable to fabrication on a chip.

S. Jahani acknowledges Zubin Jacob for discussions.

\cleardoublepage
\renewcommand{\thefigure}{S\arabic{figure}}
\renewcommand{\thesection}{S.\arabic{section}}
\renewcommand{\theequation}{S\arabic{equation}}

\setcounter{section}{0}
\setcounter{figure}{0}
\setcounter{equation}{0}

\begin{widetext}
\begin{center}

\begingroup
    \fontsize{14pt}{12pt}\selectfont
    \bf{Confining light in all-dielectric anisotropic metamaterial particles for nano-scale nonlinear optics: Supplementary Materials}
    
\endgroup
\bigskip
Saman Jahani$^1$, Joong Hwan Bahng$^1$, Arkadev Roy$^1$, Nicholas Kotov$^2$, and Alireza Marandi$^1$\\
\bigskip
$^1$Department of Electrical Engineering, 
California Institute of Technology, Pasadena, CA 91125, USA.
\\
$^2$Department of Chemical Engineering, University of Michigan, Ann Arbor, MI 48109, USA.
\\
\end{center}
{\small In this supplementary material, we report the wave equations in the spherical coordinate with radial anisotropy. We derive the analytical solutions to the electric and magnetic fields. We demonstrate the scattering by an anisotropic sphere excited by a plane-wave. We also compare full-wave simulation of a practical structure with our analytical calculations to confirm the validity of our model.
}

\section{Helmholtz equations}

In a homogeneous medium with spherical uniaxial anisotopic permittivity with the optical axis in the $r$ direction ($\bar{\bar{\varepsilon}}=[\varepsilon_{r},\varepsilon_{\bot},\varepsilon_{\bot}]$, where $\varepsilon_{\theta}=\varepsilon_{\varphi}=\varepsilon_{\bot}$), any arbitrary electromagnetic field in spherical coordinate can be constructed as a superposition of TM ($H_r=0$) and TE ($E_r=0$) modes. We can write the scalar Helmholtz equation for $E_r$ and $H_r$, and then derive the electric and magnetic fields in the $\theta$ and $\varphi$ directions from the fields in the $r$ direction. For the TE modes, the Helmhotrz equation can be written as:
\begin{align}
    (\nabla\times\nabla\times\vec{H})_r=k_0^2\varepsilon_{\bot}H_r.
\end{align}
Since $\nabla.\vec{H}=0$, the above equation is simplified to the familiar form of the wave equation  \cite{jackson_classical_1975}:
\begin{align}\label{Helmholtz:Hr}
    \nabla^2(rH_r)+k_0^2\varepsilon_{\bot}(rH_r)=0.
\end{align}
For the TM modes, we can write the scalar Helmhotz equation for $E_r$ as:
\begin{align}\label{Helmholtz:Er}
    (\nabla\times\nabla\times\vec{E})_r=k_0^2\varepsilon_rE_r. 
\end{align}

\begin{figure}
\centering
\begin{tabular}{cc}

\includegraphics[width=17cm]{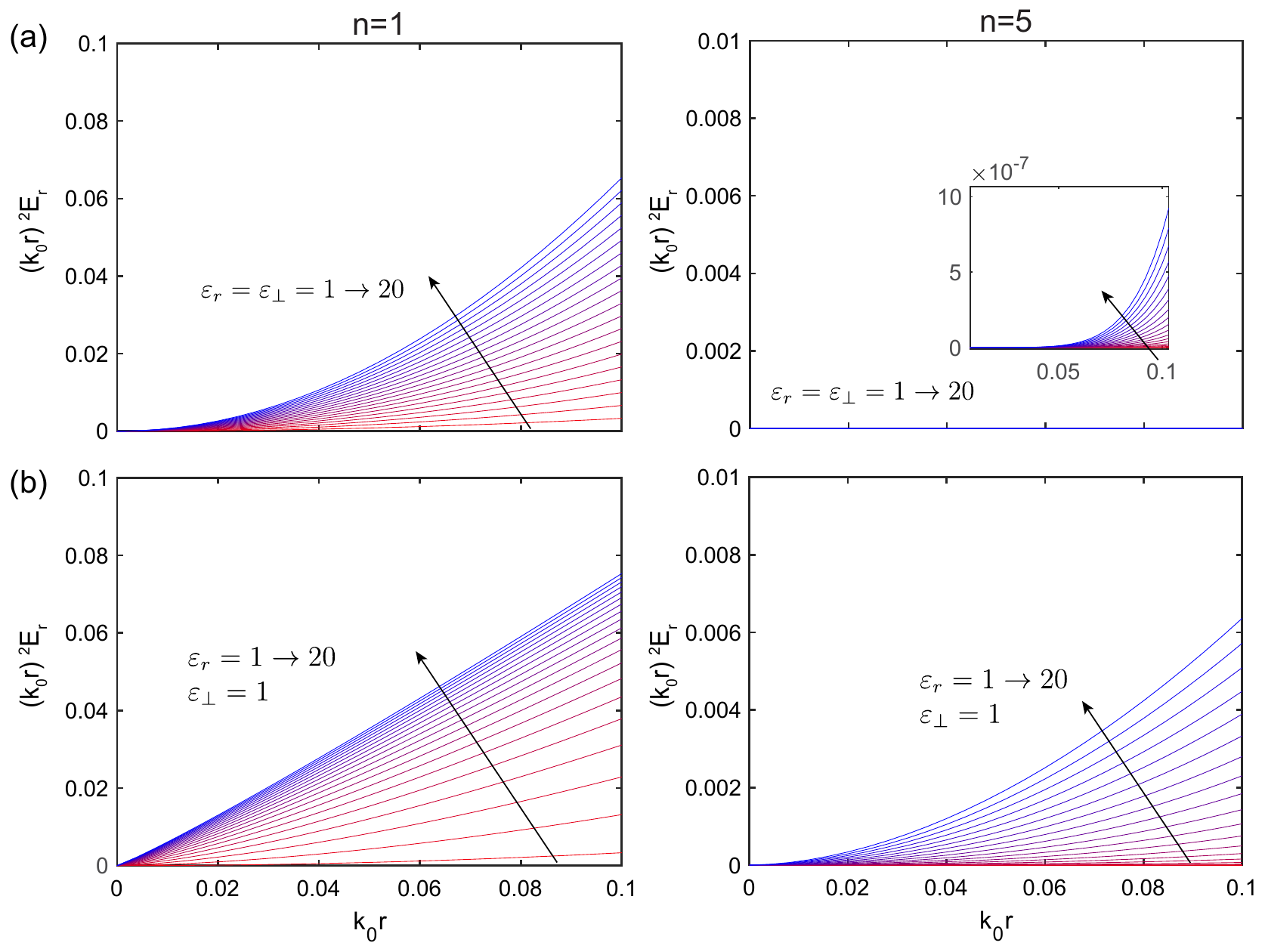}

\end{tabular}
\caption{{\bf Field distribution in an infinite homogeneous media with radial anisotropy}. Normalized electric field distribution for the first (left) and the fifth (right) electric modes as a function of permittivity. (a) Isotropic media. (b) Anisotropic media with $\varepsilon_\bot=\varepsilon_\theta=\varepsilon_\varphi=1$. $\varepsilon_\bot$ controls the momentum of spherical waves while $\varepsilon_\bot/\varepsilon_r$ changes the order of spherical Bessel waves. By increasing $\varepsilon_r$ while $\varepsilon_\bot$ is fixed, we can reduce the order without increasing the momentum. This results in an enhanced field intensity, especially in the sub-wavelength regime ($k_0r \ll 1$).   
} 
\label{fig:Anisotropic_particle_Waves_zoomed}
\end{figure}

However, since $\nabla.\vec{E}$ is not zero in anisotropic media, Eq.~\ref{Helmholtz:Er} is not simplified to the conventional form. Here, we show how we can write the Helmholtz equation for $E_r$ for the anisotropic case. The left hand side of the Eq.~\ref{Helmholtz:Er} can be written as:
\begin{align}\label{Helmholtz:Er_2}
(\nabla\times\nabla\times\vec{E})_r &= \frac{1}{r\sin{\theta}}\left[\frac{\partial}{\partial{\theta}}\left( (\nabla\times\vec{E})_{\varphi}\sin{\theta}\right) - \frac{\partial}{\partial{\varphi}} \left((\nabla\times\vec{E})_{\theta} \right) \right]  \\ \nonumber
&= \frac{1}{r\sin{\theta}} \left[\frac{\partial}{\partial{\theta}}\left(\frac{1}{r}\left(\frac{\partial}{\partial{r}}(rE_{\theta}) -\frac{\partial}{\partial\theta}E_r \right) \sin{\theta}\right)
- \frac{\partial}{\partial\varphi}\left(\frac{1}{r}\left(\frac{1}{\sin{\theta}}\frac{\partial}{\partial\varphi}E_r- \frac{\partial}{\partial{r}}(rE_{\varphi}) \right) \right) \right]  \\ \nonumber
&=-\frac{1}{r^2\sin{\theta}}\frac{\partial}{\partial\theta}\left(\sin{\theta}\frac{\partial{E_r}}{\partial\theta} \right) -\frac{1}{r^2\sin^2{\theta}}\frac{\partial^2E_{r}}{\partial\varphi^2}
+\frac{1}{r^2\sin{\theta}}\frac{\partial}{\partial\theta}\left(\sin{\theta}\frac{\partial}{\partial{r}}({E_{\theta}}) \right)
+\frac{1}{r^2\sin{\theta}}\frac{\partial}{\partial\varphi}\frac{\partial}{\partial{r}}(rE_{\varphi}) \\ \nonumber
&=-\nabla^2_{\bot}E_r +\frac{1}{r^2\sin{\theta}}\frac{\partial}{\partial\theta}\left(\sin{\theta}\frac{\partial}{\partial{r}}(r{E_{\theta}}) \right)
+\frac{1}{r^2\rm{sin}\theta}\frac{\partial}{\partial\varphi}\frac{\partial}{\partial{r}}(rE_{\varphi}),
\end{align}
where $\nabla^2_{\bot}$ is the transverse component of the Laplacian in the spherical coordinate. We can further simplify Eq.~\ref{Helmholtz:Er_2} by adding and subtracting the radial component of the Laplacian which is multiplied by $\varepsilon_r/(r\varepsilon_{\bot})$:
\begin{align}\label{Helmholtz:Er_3}
(\nabla\times\nabla\times\vec{E})_r &=-\frac{1}{r}\nabla^2_{\bot}(rE_r)-\frac{\varepsilon_r}{r\varepsilon_{\bot}}\frac{1}{r^2}\frac{\partial}{\partial{r}}\left(r^2\frac{\partial}{\partial{r}}(rE_r)\right)\\ \nonumber
&+\left[\frac{\varepsilon_r}{r\varepsilon_{\bot}}\frac{1}{r^2}\frac{\partial}{\partial{r}}\left(r^2\frac{\partial}{\partial{r}}(rE_r)\right)
+\frac{1}{r^2\sin{\theta}}\frac{\partial}{\partial\theta}\left(\sin{\theta}\frac{\partial}{\partial{r}}(r{E_{\theta}}) \right)
+\frac{1}{r^2\sin{\theta}}\frac{\partial}{\partial\varphi}\frac{\partial}{\partial{r}}(rE_{\varphi}) \right].
\end{align}
After some algebra, it is easy to show that the last term on the right side of Eq.~\ref{Helmholtz:Er_3} can be written as the divergence of the displacement current:
\begin{align}\label{Helmholtz:Er_4}
(\nabla\times\nabla\times\vec{E})_r &=\frac{1}{r} \left[-\nabla^2_{\bot}(rE_r)-\frac{\varepsilon_r}{\varepsilon_{\bot}}\frac{1}{r^2}\frac{\partial}{\partial{r}}\left(r^2\frac{\partial}{\partial{r}}(rE_r)\right)
+\frac{1}{\varepsilon_0\varepsilon_\bot}\nabla.\vec{D}+\frac{1}{\varepsilon_0\varepsilon_\bot}\frac{\partial}{\partial{r}}\left(r\nabla.\vec{D} \right) \right] \\ 
\end{align}
As $\nabla.\vec{D}=0$, the Helmholtz equation for $E_r$ can be written as:
\begin{align}\label{Helmholtz:Er_4}
\frac{\varepsilon_r}{\varepsilon_{\bot}}\frac{1}{r^2}\frac{\partial}{\partial{r}}\left(r^2\frac{\partial}{\partial{r}}(rE_r)\right) + \nabla^2_{\bot}(rE_r)
+k_0^2\varepsilon_{r}(rE_r)=0,
\end{align}
or it can be expressed as:
\begin{align}
\label{eqn:wave_eqn_momentum}
   \frac{1}{\varepsilon_{\bot}}{\pmb{p}}_r^2(rE_r)  + \frac{1}{\varepsilon_{r}r^2}\Vec{\pmb{L}}^2(rE_r)
=k_0^2(rE_r),
\end{align}
where $\hbar{\pmb{p}}_r= \frac{\hbar}{i}(\hat{r}.\Vec{\nabla})$ and $\hbar\Vec{\pmb{L}}=\frac{\hbar}{i}(\Vec{r}\times \Vec{\nabla})$ are the radial momentum and the angular momentum operators with eigenvalues of $\hbar k_r$ and $\hbar L=\hbar\sqrt{n(n+1)}$, respectively. Hence, the eigenvalue problem can be simplified to:
\begin{align}
\label{eqn:wave_eqn_momentum_eigenvaule}
   \frac{k_r^2}{\varepsilon_{\bot}}  + \frac{n(n+1)}{\varepsilon_{r}r^2} =k_0^2.
\end{align}
By increase the angular momentum mode number, the second term on the left hand side of Eq.~\ref{eqn:wave_eqn_momentum_eigenvaule} exceeds the term on the right hand side, especially when we are closer to the center. As a result, the radial momentum becomes imaginary which decays evanescently when we approach the center. This causes a weak excitation of higher order modes  in the sub-wavelength regime and a weak coupling of these modes to the far-field radiating modes \cite{jacob2006optical}.

If we rearrange the momentum as:
\begin{align}
\label{eqn:momentum}
   k_r   = \sqrt{\frac{{\varepsilon_{\bot}}}{{\varepsilon_{r}}}}\sqrt{k_0^2\varepsilon_{r}-\frac{n(n+1)}{r^2}},
\end{align}
it is seen that by controlling the anisotropy, we can  control the evanescent fields near the center \cite{jahani2014transparent}. Especially, if we increase the ratio, $\varepsilon_\bot/\varepsilon_r$, the evanescent fields and as a result, the field is enhanced in the sub-wavelength regime ($k_0r \ll 1$). Figure~\ref{fig:Anisotropic_particle_Waves_zoomed} demonstrates the fields in the sub-wavelength regime. Intuitively, it is expected to enhance the field near the center by increasing the permittivity. However, it is seen that in the anisotropic cases, even though the averaged permittivity is lower, the field enhancement is more significant. This field enhancement can be several orders of magnitude stronger for the higher order modes. 

\section{Solution to the Helmholtz equations}

\subsection{Non-magnetic anisotropic particle}

We start with the simplest particle with non-magnetic anisotropy. We can use the approach of separating the variables to find the solutions of $E_r$ and $H_r$. Eq.~\ref{Helmholtz:Hr} has the standard solution of \cite{jackson_classical_1975}:
\begin{align}\label{Hr:solution}
    rH_r(r,\theta,\varphi)=\sum_{n=0}^{\infty}\sum_{m=-n}^{n} \left[ c_n^{h}j_n(k_0\sqrt{\varepsilon_{\bot}}r) +d_n^{h}n_n(k_0\sqrt{\varepsilon_{\bot}}r) \right]P_n^{(m)}\left(\cos{\theta}\right) \left\{ \begin{matrix} \sin{(m\varphi)} \\ \cos{(m\varphi)} \end{matrix} \right\},
\end{align}
where $P_n^{(m)}$ are the Legendre Polynomials. $j_n$ and $n_n$ are the spherical Bessel and Neumann functions
defined as:
\begin{align}
    j_n(x) &= \left(\frac{\pi}{2x} \right)^{\frac{1}{2}}J_{n+\frac{1}{2}}(x) \\ \nonumber
    n_n(x) &= \left(\frac{\pi}{2x} \right)^{\frac{1}{2}}N_{n+\frac{1}{2}}(x),
\end{align}
where $J_n(x)$ and $N_n(x)$ are the $n^\text{th}$ order Bessel and Neumann functions. Sometimes, it is more convenient to write the solution as Ricatti-Bessel functions defined as \cite{kerker2013scattering, bohren2008absorption}:
\begin{align}
    \psi_n(x) &= xj_n(x)= \left(\frac{\pi{x}}{2} \right)^{\frac{1}{2}}J_{n+\frac{1}{2}}(x) \\ \nonumber
    \chi_n(x) &= -xn_n(x)=-\left(\frac{\pi{x}}{2} \right)^{\frac{1}{2}}N_{n+\frac{1}{2}}(x),
\end{align}
or as spherical Hankel function of the first kind and second kind for outward and inward radiations, respectively:
\begin{align}
    h^{(1)}_n(x) &= \xi_n(x)/x= j_n(x)+in_n(x) \\ \nonumber
    h^{(2)}_n(x) &= \zeta_n(x)/x= j_n(x)-in_n(x)
\end{align}

The angular part of the solution of Eq.~\ref{Helmholtz:Er_4} is the same as that in Eq.~\ref{Hr:solution}. However, the radial part is a bit more complicated than the standard form shown in Eq.~\ref{Hr:solution}:
\begin{align}\label{Er:solution}
    rE_r(r,\theta,\varphi)=\sum_{n=0}^{\infty}\sum_{m=-n}^{n} \left[ c_n^{e}j_{n_e}(k_0\sqrt{\varepsilon_{\bot}}r) +d_n^{e}n_{n_e}(k_0\sqrt{\varepsilon_{\bot}}r) \right]P_n^{(m)}\left(\cos{\theta}\right)\left\{ \begin{matrix} \sin{(m\varphi)} \\ \cos{(m\varphi)} \end{matrix} \right\},
\end{align}
where ${n_e}=\sqrt{\frac{\varepsilon_\bot}{\varepsilon_r}n(n+1)+\frac{1}{4}}-\frac{1}{2}$. Note that if the medium is isotropic, the solution is simplified to the standard solution as shown in Eq.~\ref{Hr:solution}.

The tangential component of the electric and magnetic fields in the spherical anisotropic medium are expressed as: 
\begin{align}\label{eq.tangential:H}
    i\omega\mu_0H_{\theta} &= \frac{1}{r}\left(\frac{1}{\sin{\theta}}\frac{\partial}{\partial\varphi}E_{r}- \frac{\partial}{\partial{r}}(rE_{\varphi}) \right) \\ \nonumber
    i\omega\mu_0H_{\varphi} &= \frac{1}{r}\left(\frac{\partial}{\partial{r}}(rE_{\theta})- \frac{\partial}{\partial{\theta}}E_{r} \right),
\end{align}
and
\begin{align}\label{eq.tangential:E}
    -i\omega\varepsilon_0\varepsilon_{\bot}E_{\theta} &= \frac{1}{r}\left(\frac{1}{\sin{\theta}}\frac{\partial}{\partial\varphi}H_{r}- \frac{\partial}{\partial{r}}(rH_{\varphi}) \right) \\ \nonumber
    -i\omega\varepsilon_0\varepsilon_{\bot}E_{\varphi} &= \frac{1}{r}\left(\frac{\partial}{\partial{r}}(rH_{\theta})- \frac{\partial}{\partial{\theta}}H_{r} \right).
\end{align}

\subsubsection{TE modes}
For TE modes, $E_r=0$, so Eq.~\ref{eq.tangential:H} is simplified to:
\begin{align}\label{eq.tangential:TE}
    i\omega\mu_0H_{\theta}^{\rm TE} &= -\frac{1}{r}\frac{\partial}{\partial{r}}(rE_{\varphi}^{\rm TE})  \\ \nonumber
    i\omega\mu_0H_{\varphi}^{\rm TE} &= \frac{1}{r}\frac{\partial}{\partial{r}}(rE_{\theta}^{\rm TE}).
\end{align}
By replacing Eq.~\ref{eq.tangential:TE} into Eq.~\ref{eq.tangential:E} and multiplying the both sides by $i\omega\mu_0r$, we obtain:
\begin{align}\label{eq.tangential:E_TE2}
    \frac{\partial^2}{\partial{r^2}}(rE_{\theta}^{\rm TE})+k_0^2\varepsilon_{\bot}rE_{\theta}^{\rm TE} &= \frac{i\omega\mu_0}{\sin{\theta}}\frac{\partial}{\partial\varphi}(rH_{r}) \\ \nonumber
    \frac{\partial^2}{\partial{r^2}}(rE_{\varphi}^{\rm TE})+k_0^2\varepsilon_{\bot}rE_{\varphi}^{\rm TE} &= - i\omega\mu_0\frac{\partial}{\partial{\theta}}(rH_{r}).
\end{align}
Since the radial part of the right-hand side of Eq.~\ref{eq.tangential:E_TE2} is a spherical Bessel function, the radial part of the left-hand side must be a spherical function too. Using the recurrence relation for spherical Bessel functions:
\begin{align}
\frac{\partial^2}{\partial{r}^2}\left( rz_n(kr)\right) + \left(k^2-\frac{n(n+1)}{r}\right)rz_n(kr)=0,    
\end{align}
where $z_n(kr)$ is a spherical Bessel, Neumann, or Hankel function, Eq.~\ref{eq.tangential:E_TE2} is simplified to:
\begin{align}\label{eq.tangential:E_TE3}
    E_{\theta}^{\rm TE} &= \frac{i\omega\mu_0}{n(n+1)\sin{\theta}}\frac{\partial}{\partial\varphi}(rH_{r}) \\ \nonumber
    E_{\varphi}^{\rm TE} &= - \frac{i\omega\mu_0}{n(n+1)}\frac{\partial}{\partial{\theta}}(rH_{r}).
\end{align}
Now if we insert Eq.~\ref{eq.tangential:E_TE3} into Eq.~\ref{eq.tangential:TE}, we can obtain the tangential component of the magnetic field:
\begin{align}\label{eq.tangential:H:TE}
    H_{\theta}^{\rm TE} &= -\frac{1}{i\omega\mu_0}\frac{1}{r}\frac{\partial}{\partial{r}}(rE_{\varphi}^{\rm TE})= \frac{1}{n(n+1)}\frac{1}{r}\frac{\partial^2}{\partial{\theta}\partial{r}}\left(r^2H_{r}\right)  \\ \nonumber
    H_{\varphi}^{\rm TE} &= \frac{1}{i\omega\mu_0}\frac{1}{r}\frac{\partial}{\partial{r}}(rE_{\theta}^{\rm TE})= \frac{1}{n(n+1)}\frac{1}{r\sin{\theta}}\frac{\partial^2}{\partial{\varphi}\partial{r}}\left(r^2H_{r}\right) .
\end{align}

\subsubsection{TM modes}
For TM modes, $H_r=0$. If we follow the same procedure that we used for TE modes, the tangential electric and magnetic fields can be expressed as:
\begin{align}\label{eq.tangential:H_TM}
    H_{\theta}^{\rm TM} &= -\frac{i\omega\varepsilon_0\varepsilon_{\bot}}{n_e(n_e+1)\sin{\theta}}\frac{\partial}{\partial\varphi}(rE_{r}) \\ \nonumber
    H_{\varphi}^{\rm TM} &= \frac{i\omega\varepsilon_0\varepsilon_{\bot}}{n_e(n_e+1)}\frac{\partial}{\partial{\theta}}(rE_{r}),
\end{align}
and
\begin{align}\label{eq.tangential:E:TM}
    E_{\theta}^{\rm TM} &= \frac{1}{i\omega\varepsilon_0\varepsilon_{\bot}}\frac{1}{r}\frac{\partial}{\partial{r}}(rH_{\varphi}^{\rm TM})= \frac{1}{n_e(n_e+1)}\frac{1}{r}\frac{\partial^2}{\partial{\theta}\partial{r}}\left(r^2E_{r}\right)  \\ \nonumber
    E_{\varphi}^{\rm TM} &= -\frac{1}{i\omega\varepsilon_0\varepsilon_{\bot}}\frac{1}{r}\frac{\partial}{\partial{r}}(rH_{\theta}^{\rm TM})= \frac{1}{n_e(n_e+1)}\frac{1}{r\sin{\theta}}\frac{\partial^2}{\partial{\varphi}\partial{r}}\left(r^2E_{r}\right) .
\end{align}

\subsection{General anisotropic particle}

For a particle with both electric and magnetic anisotropy ($\bar{\bar{\varepsilon}}=[\varepsilon_{r},\varepsilon_{\bot},\varepsilon_{\bot}]$ and $\bar{\bar{\mu}}=[\mu_{r},\mu_{\bot},\mu_{\bot}]$, where $\varepsilon_{\theta}=\varepsilon_{\varphi}=\varepsilon_{\bot}$ and $\mu_{\theta}=\mu_{\varphi}=\mu_{\bot}$), the solution for both TE and TM modes are affected by the anisotropy \cite{qiu2005field,qiu2008peculiarities, liu2015ultra}. The radial component of the electromagnetic fields are written as:
\begin{align}\label{Er and Hr:solution_both}
    rE_r(r,\theta,\varphi) &=\sum_{n=0}^{\infty}{\sum_{m=-n}^{n}{ \left[c_n^{e}j_{n_e}(k_0\sqrt{\varepsilon_{\bot}\mu_{\bot}}r) +d_n^{e}n_{n_e}(k_0\sqrt{\varepsilon_{\bot}\mu_{\bot}}r)\right] P_n^{(m)}\left(\cos{\theta}\right) \left\{ \begin{matrix} \sin{(m\varphi)} \\ \cos{(m\varphi)} \end{matrix} \right\} }}, \\ \nonumber
    rH_r(r,\theta,\varphi) &=\sum_{n=0}^{\infty}{\sum_{m=-n}^{n}{ \left[ c_n^{h}j_{n_h}(k_0\sqrt{\varepsilon_{\bot}\mu_{\bot}}r) +d_n^{h}n_{n_h}(k_0\sqrt{\varepsilon_{\bot}\mu_{\bot}}r) \right]P_n^{(m)}\left(\cos{\theta}\right) \left\{ \begin{matrix} \sin{(m\varphi)} \\ \cos{(m\varphi)} \end{matrix} \right\} }},
\end{align}
where ${n_e}=\sqrt{\frac{\varepsilon_\bot}{\varepsilon_r}n(n+1)+\frac{1}{4}}-\frac{1}{2}$ and ${n_h}=\sqrt{\frac{\mu_\bot}{\mu_r}n(n+1)+\frac{1}{4}}-\frac{1}{2}$.

The tangential components of the electric and magnetic fields for TE modes can be written as:
\begin{align}\label{eq.tangential:E_TE_both}
    E_{\theta}^{\rm TE} &= \frac{i\omega\mu_0\mu_\bot}{n_h(n_h+1)\sin{\theta}}\frac{\partial}{\partial\varphi}(rH_{r}) \\ \nonumber
    E_{\varphi}^{\rm TE} &= - \frac{i\omega\mu_0\mu_\bot}{n_h(n_h+1)}\frac{\partial}{\partial{\theta}}(rH_{r}),
\end{align}
and 
\begin{align}\label{eq.tangential:H:TE_both}
    H_{\theta}^{\rm TE} &= -\frac{1}{i\omega\mu_0\mu_\bot}\frac{1}{r}\frac{\partial}{\partial{r}}(rE_{\varphi}^{\rm TE})= \frac{1}{n_h(n_h+1)}\frac{1}{r}\frac{\partial^2}{\partial{\theta}\partial{r}}\left(r^2H_{r}\right)  \\ \nonumber
    H_{\varphi}^{\rm TE} &= \frac{1}{i\omega\mu_0\mu_\bot}\frac{1}{r}\frac{\partial}{\partial{r}}(rE_{\theta}^{\rm TE})= \frac{1}{n_h(n_h+1)}\frac{1}{r\sin{\theta}}\frac{\partial^2}{\partial{\varphi}\partial{r}}\left(r^2H_{r}\right) .
\end{align}
For the TM modes, the the tangential components are:
\begin{align}\label{eq.tangential:H_TM_both}
    H_{\theta}^{\rm TM} &= -\frac{i\omega\varepsilon_0\varepsilon_{\bot}}{n_e(n_e+1)\sin{\theta}}\frac{\partial}{\partial\varphi}(rE_{r}) \\ \nonumber
    H_{\varphi}^{\rm TM} &= \frac{i\omega\varepsilon_0\varepsilon_{\bot}}{n_e(n_e+1)}\frac{\partial}{\partial{\theta}}(rE_{r}),
\end{align}
and
\begin{align}\label{eq.tangential:E:TM_both}
    E_{\theta}^{\rm TM} &= \frac{1}{i\omega\varepsilon_0\varepsilon_{\bot}}\frac{1}{r}\frac{\partial}{\partial{r}}(rH_{\varphi}^{\rm TM})= \frac{1}{n_e(n_e+1)}\frac{1}{r}\frac{\partial^2}{\partial{\theta}\partial{r}}\left(r^2E_{r}\right)  \\ \nonumber
    E_{\varphi}^{\rm TM} &= -\frac{1}{i\omega\varepsilon_0\varepsilon_{\bot}}\frac{1}{r}\frac{\partial}{\partial{r}}(rH_{\theta}^{\rm TM})= \frac{1}{n_e(n_e+1)}\frac{1}{r\sin{\theta}}\frac{\partial^2}{\partial{\varphi}\partial{r}}\left(r^2E_{r}\right) .
\end{align}

\begin{figure}
\centering
\begin{tabular}{cc}

\includegraphics[width=15cm]{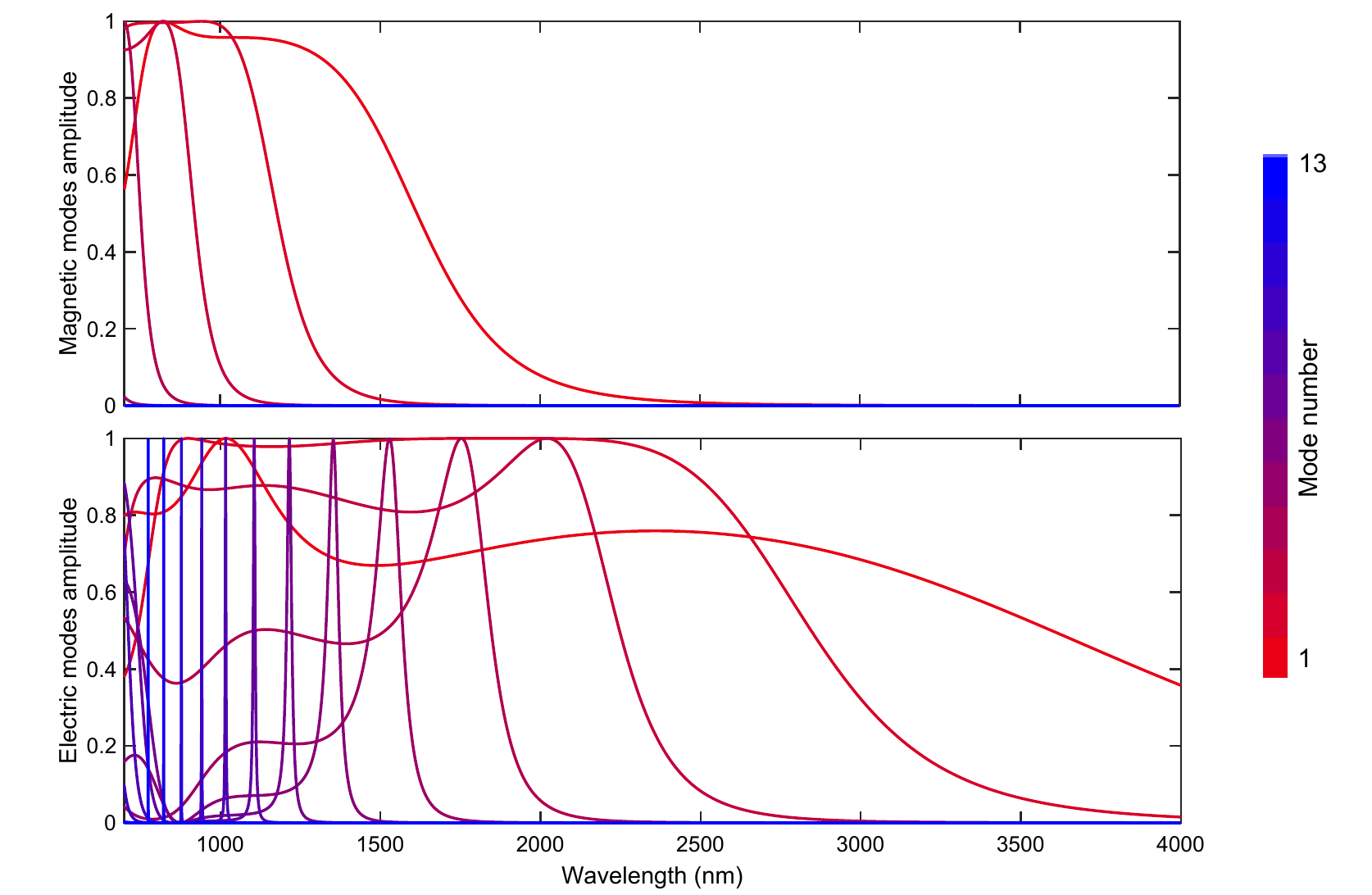}

\end{tabular}
\caption{Scattering amplitude for electric ($|a_n|^2$) and magnetic modes ($|b_n|^2$) for a low-index particle with anisotropic metamaterial shell. The parameters for the particle are the same as those in Fig.~4 in the main text.
} 
\label{fig:Anisotropic_particle_Scattering_anisotropic}
\end{figure}

\begin{figure}
\centering
\begin{tabular}{cc}

\includegraphics[width=15cm]{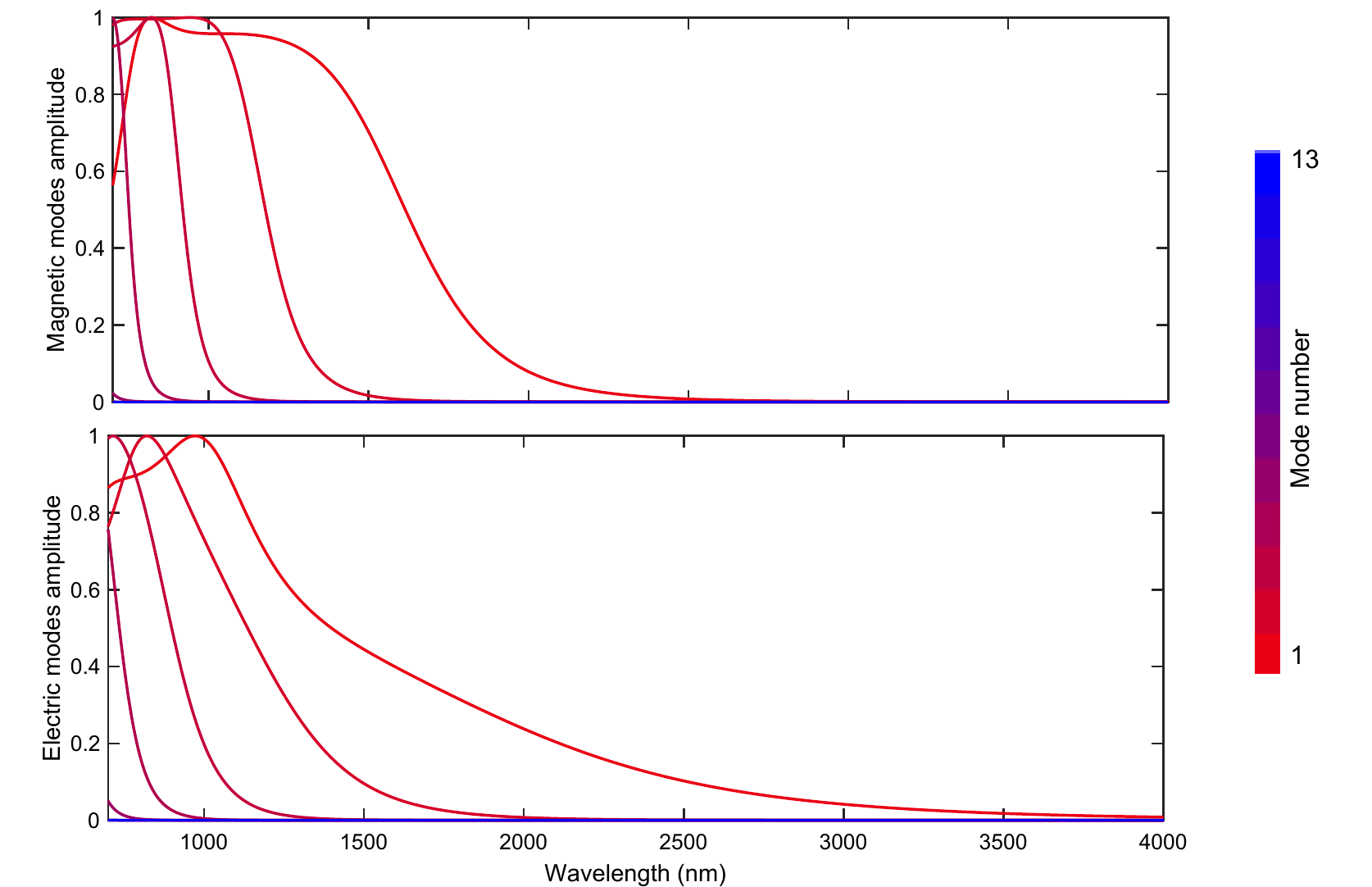}

\end{tabular}
\caption{Scattering amplitude for electric ($|a_n|^2$) and magnetic modes ($|b_n|^2$) for a the structure shown in Fig.~\ref{fig:Anisotropic_particle_Scattering_anisotropic} without the metamaterial shell.
} 
\label{fig:Anisotropic_particle_Scattering_isotropic}
\end{figure}

\begin{figure}
\centering
\begin{tabular}{cc}

\includegraphics[width=15cm]{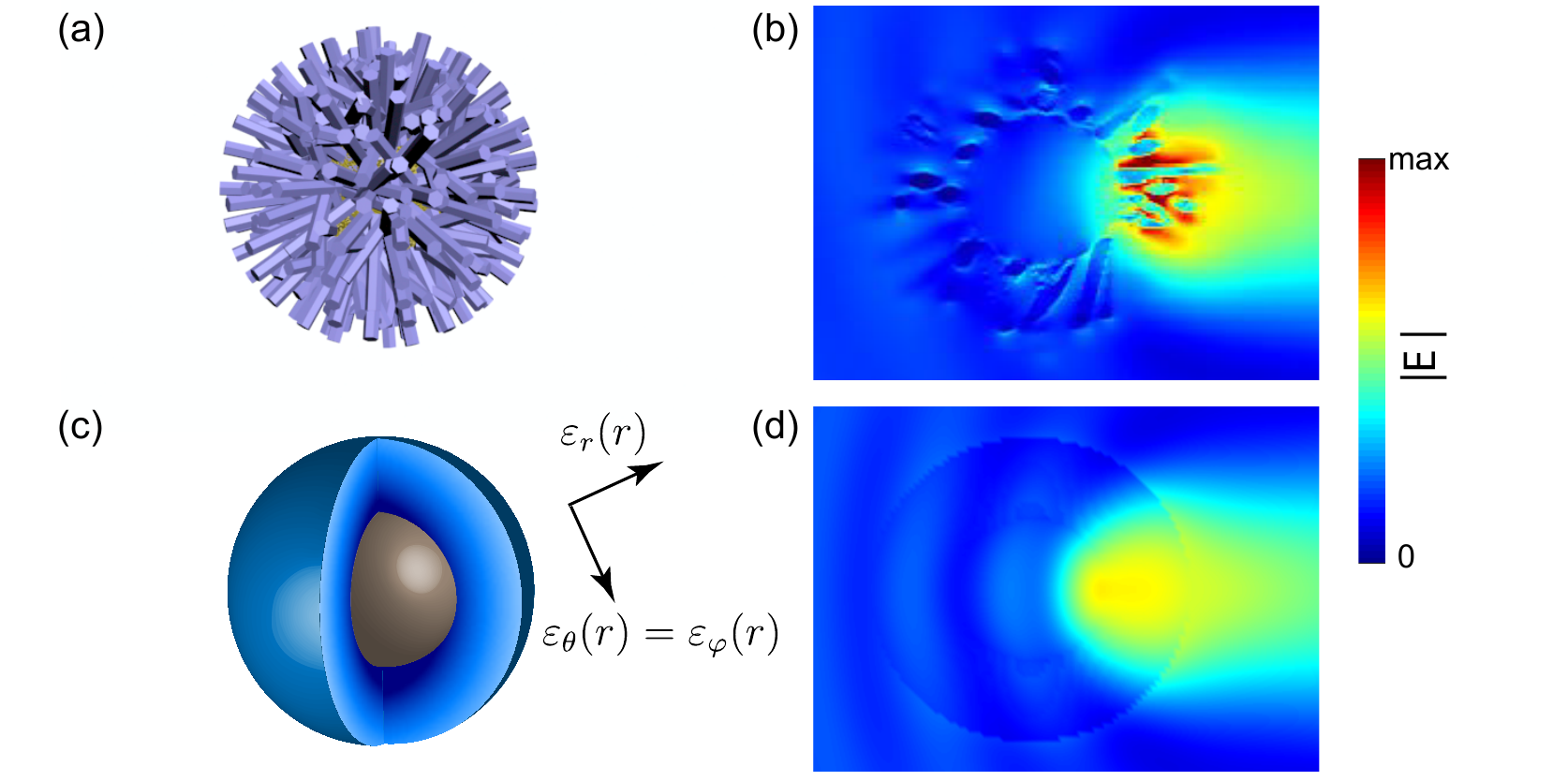}

\end{tabular}
\caption{
{\bf  Practical realization of particles with anisotropic metamaterial shell.} (a) A schematic representation of a particle with metamaterial shell. The core is composed of glass with a radius of 500 $\mu$m. The outer radius of the shell is 2.2 $\mu$m. The nanowires are composed of zinc oxide with a width of 200~nm with a filling factor of 0.4 at the core/shell interface. (b) The electric field distribution in the particle when it is excited by a plane wave propagating (wavelength: 1550 nm) to the right. (c) Effective medium modeling of the shell with anisotropic metamaterial. (d) The electric field distribution in the particle with effective medium modeling. It is seen that there is a good agreement between the full-wave numerical simulation of the real structure and the homogenized one.
} 
\label{fig:Anisotropic_particle_EMT}
\end{figure}

\section{Scattering by an anisotropic sphere}

Assuming the incident wave is a $x$ polarized plane wave travelling in the $z$ direction:
\begin{align}
    \vec{E}^i=\hat{x}E_0e^{ik_0z}=\hat{x}E_0e^{ik_0r\cos{\theta}},
\end{align}
the incident electric and magnetic fields in the spherical coordinate can be written as \cite{harrington1961time}:

\begin{align}
    E_r^i &=\cos{\varphi}\sin{\theta}E_x^i=\frac{E_0}{k_0^2r^2}\cos{\varphi}\sum_{n=1} i^{(n+1)}(2n+1)\psi_n(k_0r)P_n^{(1)}(\cos{\theta}) \\ \nonumber
    H_r^i &=\sin{\varphi}\sin{\theta}\frac{E_x^i}{\eta}=\frac{E_0}{\eta k_0^2r^2}\sin{\varphi}\sum_{n=1} i^{(n+1)}(2n+1)\psi_n(k_0r)P_n^{(1)}(\cos{\theta}),
\end{align}
where $E_0$ is the incident electric field amplitude and $\eta$ is the free-space impedance. Because of the interaction between the incident field and the particle, light is scattered. Since the scattered fields have to vanish in the infinity the scattered light is expressed as:
\begin{align}
    E_r^s &=-\frac{E_0}{k_0^2r^2}\cos{\varphi}\sum_{n=1} i^{(n+1)}(2n+1)a_n\xi_n(k_0r)P_n^{(1)}(\cos{\theta}) \\ \nonumber
    H_r^s &=-\frac{E_0}{\eta k_0^2r^2}\sin{\varphi}\sum_{n=1} i^{(n+1)}(2n+1)b_n\xi_n(k_0r)P_n^{(1)}(\cos{\theta}).
\end{align}
The fields inside the sphere have to vanish at the origin. Hence they can be expressed as:
\begin{align}
    E_r^r &=\frac{E_0}{k_0^2r^2}\cos{\varphi}\sum_{n=1} i^{(n+1)}(2n+1)c_n\psi_{n_e}(k_0\sqrt{\varepsilon_{\bot}\mu_{\bot}}r)P_n^{(1)}(\cos{\theta}) \\ \nonumber
    H_r^r &=\frac{E_0}{\eta k_0^2r^2}\sin{\varphi}\sum_{n=1} i^{(n+1)}(2n+1)d_n\psi_{n_h}(k_0\sqrt{\varepsilon_{\bot}\mu_{\bot}}r)P_n^{(1)}(\cos{\theta}).
\end{align}
By applying the boundary conditions at the particle interfaces:
\begin{align}
    E_{\theta}^r(k_0\sqrt{\varepsilon_{\bot}\mu_{\bot}}R)=E_{\theta}^i(k_0 R)+E_{\theta}^s(k_0 R) \\ \nonumber
    H_{\theta}^r(k_0\sqrt{\varepsilon_{\bot}\mu_{\bot}}R)=H_{\theta}^i(k_0 R)+H_{\theta}^s(k_0 R),
\end{align}
we can find $a_n$ and $b_n$, which are electric and magnetic Mie scattering coefficients, respectively. The total scattering and extinction cross-sections can be expressed as \cite{kerker2013scattering}:
\begin{align}
    C_{\rm sca} &=\frac{2\pi}{k_0^2}\sum_{n=1}^{\infty}(2n+1)\left( \mid a_n \mid^2 + \mid b_n \mid^2\right),\\ \nonumber
    C_{\rm ext} &=\frac{2\pi}{k_0^2}\sum_{n=1}^{\infty}(2n+1){\rm Re}\{a_n+b_n\}.
\end{align}

Figures~\ref{fig:Anisotropic_particle_Scattering_anisotropic} and \ref{fig:Anisotropic_particle_Scattering_isotropic} show the scattering amplitudes for the electric and magnetic modes with and without the anisotropic metamaterial shell, respectively. The parameters are the same as those in Fig.~4 in the main text. Since the magnetic modes do not feel the anisotropy, the scattering coefficients for these modes are not affected by the presence of the metamaterial shell. However, as seen in Fig.~\ref{fig:Anisotropic_particle_Scattering_anisotropic}, the anisotropic metamaterial shell significantly enhances the response to the electric modes and as a result, the excitation of higher order modes. 

\section{Practical realization}

In the last few years, several approaches have been proposed to experimentally realize particles with radial anisotropic metamaterial shell \cite{palmer2020highly, liu2019direct}. We have recently proposed and demonstrated particles with metamaterial shell in colloidal platform (Fig.~\ref{fig:Anisotropic_particle_EMT}(a)) \cite{bahng2020mie}. They are composed of a low-index nanoparticle covered by a shell of nanowires with ability to engineer and tune their optical properties. Since the feature size of the nanowires is in sub-wavelength regime, effective medium theory can be applied to homogenize the shell. We have used  Maxwell-Garnett effective medium theory \cite{jahani2014transparent} to model the nanowires with an all-dielectric metamaterial representing radial anisotropy. Since the the filling factor reduces as the radius increases, the metamaterial shell also demonstrates a graded-index profile \cite{bahng2020mie}. A comparison between the FDTD simulation \cite{lumerical_fdtd_solutions} of the practical structure and analytical calculation of the field distribution in the homogenized structure demonstrates the success of modeling of the nanowires with radial anisotropic metamaterial (Fig.~\ref{fig:Anisotropic_particle_EMT}). However, when the refractive index of the nanowires increases, and the Mie modes in the nanowires are excited, the effective medium theory needs to be modified  \cite{schuller2007dielectric, rybin2015phase}.

\end{widetext}

\bibliography{references}
\end{document}